\documentclass[letterpaper, 10 pt, conference]{ieeeconf}  

\IEEEoverridecommandlockouts                              

\usepackage{cite}
\usepackage{amssymb,amsfonts}
\usepackage{algorithm}
\usepackage{algpseudocode}
\usepackage{textcomp}
\usepackage{xcolor}
\usepackage{hyperref}       
\usepackage{url}            
\usepackage{caption}
\usepackage{tabularx}
\usepackage{booktabs}
\usepackage{tfrupee}

\usepackage{xcolor,soul,framed} 

\colorlet{shadecolor}{yellow}
\usepackage[pdftex]{graphicx}
\graphicspath{{../pdf/}{../jpeg/}}
\DeclareGraphicsExtensions{.pdf,.jpeg,.png}

\usepackage[cmex10]{amsmath}
\usepackage{array}
\usepackage{mdwmath}
\usepackage{mdwtab}
\usepackage{eqparbox}
\usepackage{url}
\usepackage{booktabs}
\usepackage{multirow}

\def\BibTeX{{\rm B\kern-.05em{\sc i\kern-.025em b}\kern-.08em
    T\kern-.1667em\lower.7ex\hbox{E}\kern-.125emX}}

\title{\LARGE \bf
Multimodal Appearance-based Gaze-Controlled Virtual Keyboard with Synchronous–Asynchronous Interaction for Low-Resource Settings
}

\author{Yogesh Kumar Meena$^{*1}$ Manish Salvi$^{1}$
\thanks{$^{1}$Yogesh Kumar Meena and Manish Salvi are with Human-AI Interaction (HAIx) Lab, IIT Gandhinagar, India. 
        {\tt\small yk.meena@iitgn.ac.in}}%
}

\begin{document}

\maketitle
\thispagestyle{empty}
\pagestyle{empty}

\begin{abstract}

Over the past decade, the demand for communication devices has increased among individuals with mobility and speech impairments. Eye-gaze tracking has emerged as a promising solution for hands-free communication; however, traditional appearance-based interfaces often face challenges such as accuracy issues, involuntary eye movements, and difficulties with extensive command sets. This work presents a multimodal appearance-based gaze-controlled virtual keyboard that utilises deep learning in conjunction with standard camera hardware, incorporating both synchronous and asynchronous modes for command selection. The virtual keyboard application supports menu-based selection with nine commands, enabling users to spell and type up to 56 English characters—including uppercase and lowercase letters, punctuation, and a delete function for corrections. The proposed system was evaluated with twenty able-bodied participants who completed specially designed typing tasks using three input modalities: (i) a mouse, (ii) an eye-tracker, and (iii) an unmodified webcam. Typing performance was measured in terms of speed and information transfer rate (ITR) at both command and letter levels. Average typing speeds were $18.3\pm5.31$ letters/min (mouse), $12.60\pm2.99$ letters/min (eye-tracker, synchronous), $10.94\pm1.89$ letters/min (webcam, synchronous), $11.15\pm2.90$ letters/min (eye-tracker, asynchronous), and $7.86\pm1.69$ letters/min (webcam, asynchronous). ITRs were approximately $80.29\pm15.72$ bits/min (command level) and $63.56\pm11$ bits/min (letter level) with webcam in synchronous mode. The system demonstrated good usability and low workload with webcam input, highlighting its user-centred design and promise as an accessible communication tool in low-resource settings.

\end{abstract}

\section{INTRODUCTION}

Information and communication technologies (ICTs) have become an integral part of daily life; however, individuals with disabilities often face challenges in effectively navigating and using these applications \cite{21}. Approximately 15\% of the global population has a disability, with 80\% of these individuals living in low-resource settings, where they frequently experience significant social isolation \cite{22}. Over the past decade, research has developed assistive technologies for alternative communication for those with severe motor and speech disabilities. These technologies utilise a range of methods, including electroencephalography (EEG), electromyography (EMG), and electro-oculography (EOG) \cite{1,2,3}. While EEG-based brain-computer interfaces (BCIs) show promise in assisting individuals with neuromuscular disorders, the requirement for constant device attachment can be uncomfortable. In this context, eye-tracking emerges as a more user-friendly and portable input method compared to existing technologies.

Integrating eye-tracking devices with virtual keyboards offers a promising communication alternative for those with severe speech and motor impairments~\cite{4,5,7,8,9,23,24} that can also be easily integrated with other methods like EEG-based BCIs to create hybrid alternatives for communication~\cite{6,30,34}. However, traditional devices can be uncomfortable or expensive, limiting accessibility for individuals with disabilities living in low-resource settings, such as developing countries with low per capita income. To overcome these challenges, there is a need for solutions that do not require additional hardware and can leverage existing infrastructure. Advances in appearance-based deep learning models for gaze estimation~\cite{24}~\cite{11}, along with open-source datasets like GazeCapture~\cite{12}, MPIIGaze~\cite{13}, and Gaze360~\cite{10}, provide effective solutions to these issues without the necessity for extra hardware.



Appearance-based gaze estimation using deep learning has shown promising results for gaze estimation~\cite{11}. However, its implementation in real-time applications, such as virtual keyboards, is still limited compared to eye-tracking and BCI-based virtual keyboard solutions for users with disabilities. To better support low-resource settings, it is essential to develop an interface compatible with everyday devices, such as laptops using built-in webcams for gaze estimation. Despite the challenges associated with low-resolution cameras, several studies have reported accuracy rates up to 98.77\% in gaze tracking with unmodified webcams\cite{14,17,18,19}. Zhang et al. \cite{15} and Huang et al. \cite{16} demonstrated effective control of virtual keyboards through CNN-based gaze prediction. Their work highlights the potential of using unmodified, low-resolution cameras for gaze estimation. However, it is essential to have substantial task-specific eye image datasets to ensure accuracy. To develop a real-time eye-tracking system, it is necessary to create a precise appearance model based on unmodified cameras and directly compare it with existing commercial eye-tracking devices~\cite{14,17,18,19,15,16}. The deep learning based gaze estimation model should take into account different interaction methods, such as synchronous and asynchronous interaction, to address accuracy issues, involuntary eye movements, and challenges associated with extensive command sets.

The main contributions of this work are as follows: 1) propose a gaze estimation model for eye direction classification using deep learning. 2) introduce both synchronous and asynchronous methods for selecting commands in webcam-based gaze interaction. 3) design a multimodal, menu-based virtual keyboard application with nine commands, allowing users to spell and type up to 56 English characters, including uppercase and lowercase letters, punctuation, and a delete function for corrections. 4) provide a benchmark with beginner users comparing traditional eye-tracking v/s webcam-based gaze control (in both synchronous and asynchronous selection modes) and mouse input as a baseline modality.

\section{Methods} \label{subsec:Methods}



\subsection {Dataset}

The dataset comprises 20000 synthetic eye images depicting gazes directed toward nine different positions: north-west, north, north-east, west, centre, east, south-west, south, and south-east. These images were generated using UnityEyes software \cite{24}. The labels in the dataset range from 1 to 9, each corresponding to one of the nine gaze directions. To enhance the model's robustness and generalisation capability, data augmentation techniques were applied. Specifically, the ImageDataGenerator from the Keras library was utilised to perform this augmentation. The augmentation included a rotation range of 5 degrees and both width and height shift ranges of 0.15, introducing variability into the dataset. We divided the dataset into an 80\% training set and a 20\% validation set. To tailor the model for individual users, we collected 2,700 calibration images—300 eye images for each direction—during a user calibration session. A separate model is fine-tuned for each participant using their own calibration data, enhancing gaze estimation accuracy. This user-specific training is essential for improving performance in eye-gaze-based typing tasks. The calibration data was divided into 75\% training and 25\% validation data.



\subsection {CNN architecture and hyperparameter tuning}

Table \ref{tab:model_summary} presents a detailed architecture of the proposed convolutional neural network (CNN) model, outlining the number and types of layers, as well as the output size and parameters. Our model is composed of four convolutional layers, each followed by a max pooling layer. At the end of the network, there are two dense layers. To prevent overfitting, three dropout layers are strategically placed throughout the model architecture. The final layer of the CNN uses the softmax activation function for classification. The input images are grayscale and sized at 100x100 pixels. We utilised the Adam optimiser and categorical cross-entropy as the loss function.


\begin{table}[]
\centering
\caption{The proposed CNN model consists of layers with
different types, kernel sizes, output sizes, and parameters.}
\begin{tabular}{lll}

\hline
\\
\textbf{Layer (type)} & \textbf{Output Shape} & \textbf{Param \#} \\  \hline \\
conv2d (Conv2D) & (16000, 90, 90, 16) & 1952 \\ 
max\_pooling2d (MaxPooling2D) & (16000, 30, 30, 16) & - \\ 
conv2d\_1 (Conv2D) & (16000, 24, 24, 32) & 25120 \\ 

max\_pooling2d\_1 (MaxPooling2D) & (16000, 8, 8, 32) & - \\ 
dropout (Dropout) & (16000, 8, 8, 32) & - \\ 

conv2d\_2 (Conv2D) & (16000, 8, 8, 64) & 51264 \\ 

max\_pooling2d\_2 (MaxPooling2D) & (16000, 4, 4, 64) & - \\ 

dropout\_1 (Dropout) & (16000, 4, 4, 64) & - \\ 

conv2d\_3 (Conv2D) & (16000, 4, 4, 128) & 73856 \\ 
max\_pooling2d\_3 (MaxPooling2D) & (16000, 2, 2, 128) & - \\ 

flatten (Flatten) & (16000, 512) & - \\ 
dense (Dense) & (16000, 250) & 128250 \\ 
dropout\_2 (Dropout) & (16000, 250) & - \\ 
dense\_1 (Dense) & (16000, 9) & 2259 \\ \hline \\
Total params: 282,701 \\
Trainable params: 282,701 \\
Non-trainable params: 0 \\ 
\hline
\end{tabular}
\label{tab:model_summary}
\end{table}

The model's capabilities can be better understood through hyperparameter tuning. We examined several hyperparameters, including the number of epochs, learning rate, batch size, and dropout rate. The number of epochs ranged from 10 to 100, with increments of 10 epochs. The batch size varied from 16 to 512, increasing in increments of 16. The learning rate was tested within a range of 0.0001 to 0.1, with a gap of 0.1 between values. Different combinations of dropout rates were applied for various layers, specifically using rates of 0.2, 0.5, and 0.9. For the model trained on synthetic images, the optimal hyperparameters were identified as a batch size of 32, a learning rate of 0.001, a dropout rate of 0.2 for the first two dropout layers, and a dropout rate of 0.5 for the last layer. The learning rate of 0.001 proved to be the most effective for our model. When fine-tuning the model on real eye images, a batch size of 128 and training for 70 epochs produced the best results.


 \subsection{Command selection with the eye-tracker}

The virtual keyboard application utilises both asynchronous and synchronous command selection methods that leverage eye-tracking technology, building on our previous work~\cite{9}. In the asynchronous selection mode, users must focus on a specific target (a box) for a predetermined duration, known as the dwell time, denoted as $\Delta t_{1}$ ms. If the user's gaze remains fixed on the same item for at least $\Delta t_{1}$ ms, that item is selected. However, if the gaze shifts to another item before this time elapses, the timer resets. In the synchronous mode, users can look anywhere, but the target that accumulates the highest weight during a defined period, referred to as the trial period ($\Delta t_{2}$ ms), will be selected. The weight assigned to a target increases linearly over time as users search for it. By the end of the trial period, the target usually has enough weight to make it more likely to be selected. The command selection process involves calculating the Euclidean distance between the centre of the target (one of the nine commands in our graphical user interface) and the gaze coordinates received from the eye-tracking device.


\begin{algorithm}[]
\caption{Command selection -- asynchronous mode}
\label{alg:async}
\begin{algorithmic}[1]
\State $\alpha \gets 6$
\While{\text{(true)}}

    \For{$t \gets 0$ to $\Delta t$}
        \State $\text{Lp} \gets \text{argmax}_{1 \leq i \leq N} (\beta_1^i t)$
        \State $\text{Rp} \gets \text{argmax}_{1 \leq i \leq N} (\beta_2^i t)$

        \If{$\text{Lp = Rp}$ and $\text{Lp} \neq 4$ (center)}
            \State $\text{selected}_t \gets \text{None}$
        \Else
            \State $\text{selected}_t \gets \text{Rp}$
           
        \EndIf
        \If{ $\text{selected}_t$ ==  $\text{selected}_{t-1}$}
            \State $\delta = \delta + 1$
        \Else
            \State $\delta = 0$
        \EndIf


        \EndFor

        \If{$\delta \geq \Delta t_{1}$}
            \State \textit{Run command}$(\text{selected}_s)$
            \State $\delta = 0$
        \Else
            \State $\text{selected}_s \gets \text{None}$
        \EndIf
\EndWhile
\end{algorithmic}
\end{algorithm}

 \subsection{Command selection with webcam}
 
Similar to command selection using eye-tracking technology, we proposed both asynchronous and synchronous methods for command selection with webcams. Instead of calculating the exact point of gaze and then determining the nearest target, our approach predicts where the user is looking using an appearance-based CNN model. The webcam continuously captures images of the user and crops the face from each frame. Using the Haar cascade algorithm~\cite{28}, we identify the left and right eyes from the extracted face image. Like eye-trackers, our webcam system selects commands based on two selection modes. However, rather than regressing the point of gaze, it classifies different targets (commands). Detailed explanations of the asynchronous and synchronous command selection methods are provided in Algorithms 1 and 2, respectively. Based on testing, we set a threshold of 6 ($\alpha \gets 6$), which corresponds to 60\% of $\Delta t_{1}$ and $\Delta t_{2}$. The frames of the left and right eyes are designated as $\text{Lp}$ and $\text{Rp}$. The total number of commands available in the system is denoted as \textit{N}. A command selection method is required to select any command from $\{1,\ldots, N\}$. For each iteration $t$, the selected command is denoted by $\text{selected}_t$ where $1 \leq \text{selected}_t \leq N$.

\begin{algorithm}[]
\caption{Command selection -- synchronous mode}
\label{alg:sync}
\begin{algorithmic}[1]
\State $\alpha \gets 6$
\While{\text{(true)}}

    \State $\text{Weights(i)} \gets 0.0, \forall i \in \{1..N\}$

    \For{$t \gets 0$ to $\Delta t_{2}$}
        \State $\text{Lp} \gets \text{argmax}_{1 \leq i \leq N} (\beta_1^i t)$
        \State $\text{Rp} \gets \text{argmax}_{1 \leq i \leq N} (\beta_2^i t)$

        \If{$\text{Lp = Rp}$ and $\text{Lp} \neq 4$ (center)}
            \State $\text{selected}_t \gets \text{None}$
        \Else
            \State $\text{selected}_t \gets \text{Rp}$
            \State $\text{Weights}(\text{selected}_t)\gets\text{Weights}(\text{selected}_t)+\sqrt{t} $
        \EndIf


        \EndFor

        \State $\text{selected}_s \gets \text{argmax}_{1 \leq i \leq N}(\text{Weights})$
        \State $P(\text{selected}_s) \gets \max(\text{Weights}) \frac{1}{N} \sum_{j=1}^{N} \text{Weights}(j)$

        \If{$P(\text{selected}_s) \geq \alpha$}
            \State \textit{Run command}$(\text{selected}_s)$
        \Else
            \State $\text{selected}_s \gets \text{None}$
        \EndIf
\EndWhile
\end{algorithmic}
\end{algorithm}

\begin{figure*}[ht!]
  \includegraphics[width=0.53\textwidth]{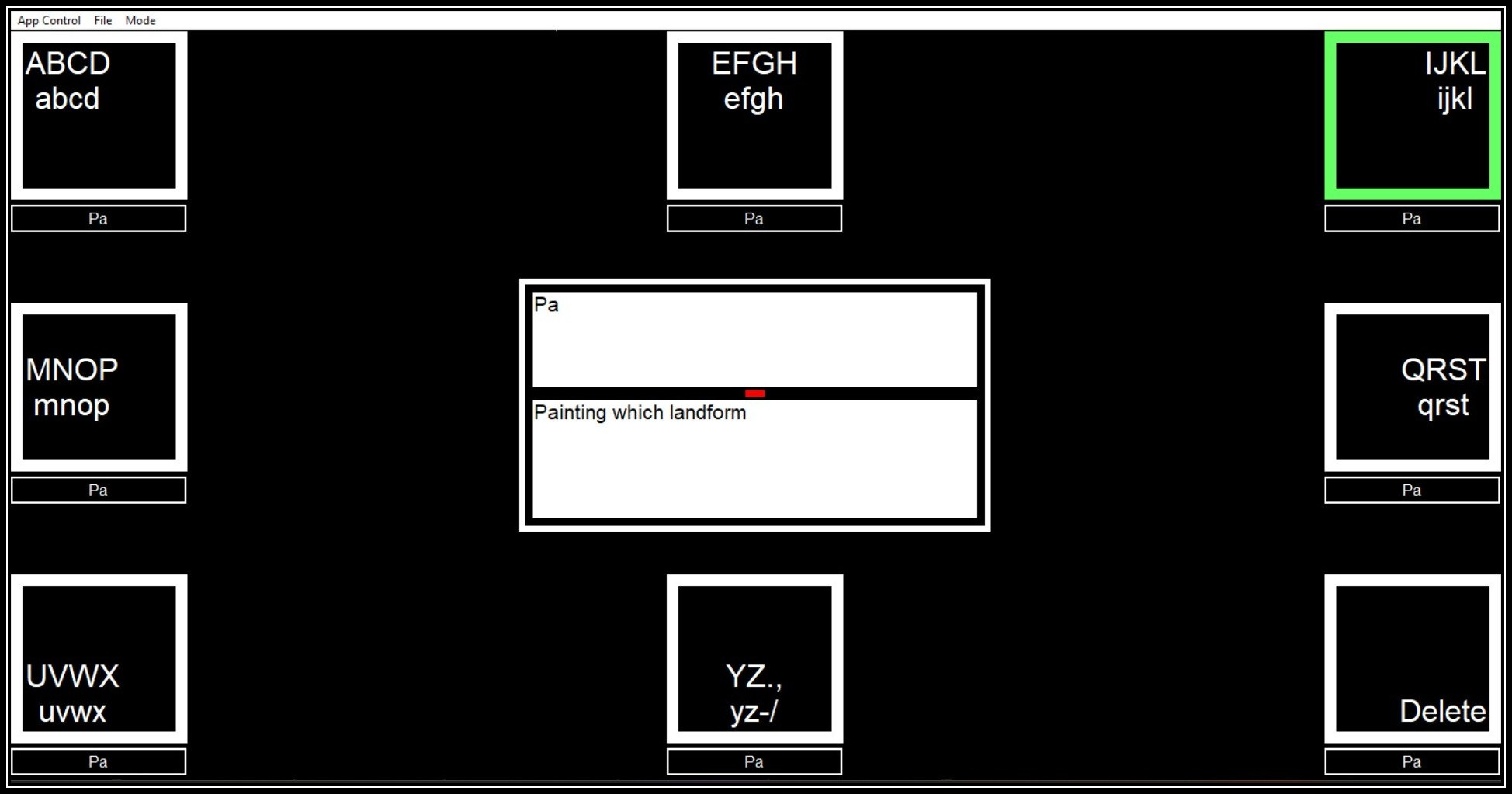}
  \hfill
  \includegraphics[width=0.44\textwidth]{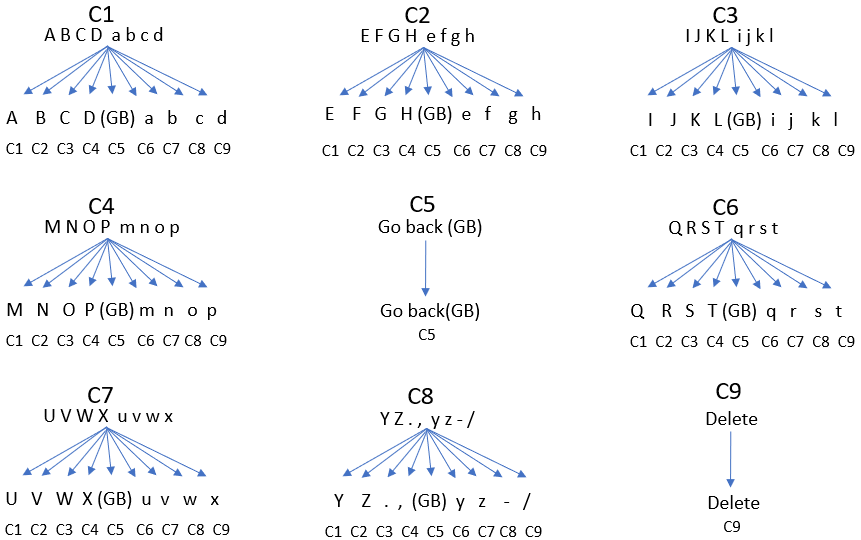}
  \caption{Graphical user interface (GUI) of proposed virtual keyboard application in level one when command three is selected (left), tree structure of all the commands present in the virtual keyboard (right).}
  \label{fig:second}
\end{figure*}



\section{System Overview} \label{subsec:System Overview}



The developed graphical user interface (GUI) features a tree-based virtual keyboard that allows users to spell and type 56 different English language characters, including capital letters, lowercase letters, and punctuation marks. The keyboard is organised into two levels, each containing nine commands positioned at north-west, north, north-east, west, east, south-west, south, south-east, and centre. At the centre of the interface, there are two text boxes and a small box in between. The first text box displays the text typed by the user, while the second text box shows the text that the user needs to type. The small box in the centre serves as a ``go back" command, enabling users to return from level 2 to level 1 after they gaze at it, in case they select an incorrect command and enter level 2 unintentionally.

In the Virtual keyboard GUI, the letters in the commands are intentionally oriented in various directions to create gaps between adjacent commands, Fig.~\ref{fig:second} (left). This design enhances the model's ability to classify eye images by introducing variations due to these gaps. Our virtual keyboard features commands C1, C2, C3, C4, C6, C7, and C8, each containing eight characters, which may include uppercase letters, lowercase letters, and punctuation marks. The command C7 incorporates punctuation marks, a space represented by a hyphen, and a ``/”. Commands C5 and C9 serve specific correction functions: ``go-back'' and ``delete'', respectively. For all other commands, once a user selects one based on the command selection mode, level 2 is activated, as illustrated in Fig. \ref{fig:second}.


When a user views any command in level 1 or level 2, the button border changes colour from white to green as visual feedback. This indicates to the user which command they are focusing on and helps prevent the selection of the wrong command. Once the user selects a letter in level 2, it is added to the text box, and audio feedback is provided that announces the chosen letter. This notification helps the user know that the current letter has been added to the text box, allowing them to focus on the next desired letter. Additionally, a rectangle-shaped box is located below each of the eight commands (except for C5, which serves as the ``go back" function), displaying the last five letters the user has typed. This feature eliminates the need for the user to constantly look at the text box in the centre. While selecting a command, the user can glance at the box below to think about the next letter they want to choose. The last five letters are updated every time a letter is added or deleted.


 \section{Experimental Protocol} \label{subsec:Experimental Protocol}

 \subsection{Participants}

Twenty consenting healthy males (eighteen) and females (two) participated in this study. They were in the age range of 22-27 years ($24.56\pm1.25$). One participant required vision correction. None of the participants had prior experience using an eye-tracker or a webcam-based eye-tracking model. Participants were informed about the purpose and nature of the study. Each participant was compensated with \rupee100 ($\approx $ \$1.21 ) for their involvement. We adhered to an informed consent procedure approved by the Institute Ethical Committee (IEC) while conducting the experiments.




\subsection{Design and operational procedure}

The experiment protocol outlined in our study involves typing a 23-character sentence: ``Painting which landform," under each experimental condition. A total of 46 commands are required to complete the experiment without any errors. This sentence was specifically designed so that each command has an equal chance of being selected, thereby minimising bias. The equiprobable sentence was crafted to evaluate the efficiency of all commands within the GUI. Although this sentence may not reflect typical real-world sentence patterns, as it contains certain letters more frequently than others, previous research has successfully employed this strategy to effectively assess GUI performance~\cite{8,9}.


Three different input modalities were utilised for the experiment: a mouse, an eye-tracker device \cite{9}, and a laptop webcam (specifically, the Asus TUF F15 HD 720p CMOS module Web Camera). With the mouse, users simply completed the task by clicking commands on a virtual keyboard. For the eye-tracking modality, each user had to undergo a calibration process using the calibration window provided by the eye-tracker device SDK. After calibration, they typed a given sentence using their gaze in both synchronous and asynchronous modes. Similarly, a nine-point calibration window was designed for the webcam, allowing users to type the same sentence using a camera-based gaze estimation model in both modes as well.

The calibration process was conducted using a webcam at 30 frames per second, similar to standard eye-tracking devices, and under typical room lighting conditions. User calibration is a one-time process that takes approximately 3 to 4 minutes. For subsequent use, the system loads the fine-tuned model, and users only need to align their head using a head pose window before beginning to type. During the eye-tracker and webcam experiments, participants were advised to minimise head and body movements as much as possible. In line with the work of Meena et al.~\cite{8,9}, the performance of the system was evaluated using measures of ${ITR}_{com}$, ${ITR}_{letter}$, and typing speed.


\section{Results} \label{subsec:Results}

\subsection{Eye direction classification}

\begin{figure*}[]
  \centering
  \includegraphics[width=0.42\textwidth]{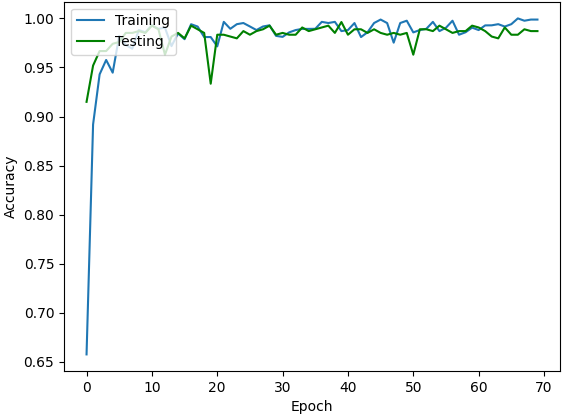}
  \includegraphics[width=0.37\textwidth]{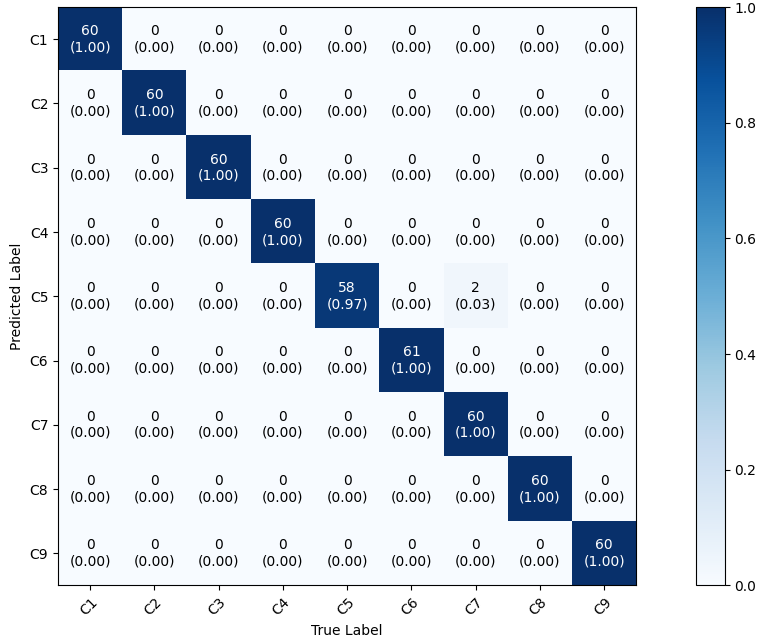}
  \caption{Training and testing accuracy's of proposed fine-tuned model over several epochs (left). Confusion matrix showing true class (command c1-c9) vs predicted class (command c1-c9) after fine-tuning (right).}
  \label{fig:accuracy}
\end{figure*}

The proposed CNN model achieved a training accuracy of 97.43\% and a validation accuracy of 98.75\% after being trained for 70 epochs. Following fine-tuning with calibration data that included 2,700 eye images (300 images for each direction) for an additional 70 epochs, the CNN model outperformed existing appearance-based gaze estimation models, as detailed in Table ~\ref{tab:first}. It achieved an average test accuracy of $99.64\pm0.3\%$ while utilising a comparatively smaller number of images. Within just 10 epochs, our proposed CNN model surpassed 95\% training and testing accuracy, as illustrated in Fig ~\ref{fig:accuracy} (left). Furthermore, Fig ~\ref{fig:accuracy} (right) illustrates the model's efficiency in performing gaze estimation using a confusion matrix.



\subsection{Typing performance}

Table \ref{tab:second} presents performance results for different conditions (Mouse, Eye-Tracker Synchronous (Sy), Webcam Synchronous (Sy), Eye-Tracker Asynchronous (As), and Webcam Asynchronous (As)) in a typing task. The mouse baseline achieved an average speed of ${18.3 \pm 5.31}$ letters/min. Moving to eye-tracking devices, the best-performing eye-tracker with a synchronous mode achieved ${12.60 \pm 2.29}$ letters/min, while the asynchronous mode recorded ${11.15 \pm 2.90}$ letters/min. Webcam conditions showed the best performance with a synchronous mode at ${10.95 \pm 1.89}$ letters/min, decreasing to ${7.86 \pm 1.69}$ letters/min with an asynchronous mode. The performance pattern, measured in 
${ITR}_{letter}$ and ${ITR}_{com}$, were similar to speed across conditions, with the mouse having the highest average ${ITR}_{letter}$ ${106.28 \pm 30.85}$ bits/min and ${ITR}_{com}$ ${118.45 \pm 32.89}$ bits/min. A decline in ${ITR}_{letter}$ and ${ITR}_{com}$ was measured as we moved towards eye-tracker and webcam conditions. We achieved ${ITR}_{letter}$ for eye-tracker synchronous (Sy), eye-tracker asynchronous (As), webcam synchronous(Sy) and webcam asynchronous (As) as  ${88.66 \pm 11.41}$, ${84.07 \pm 17.30}$, ${80.29 \pm 15.72}$ and ${71.62 \pm 22.33}$ respectively. Additionally, ${ITR}_{com}$ was found to be ${73.18 \pm 13.28}$, ${64.79 \pm 16.85}$, ${63.56 \pm 11}$ and ${45.61 \pm 9.81}$ respectively. We found that a webcam with a synchronous mode leads to a faster speed than a webcam with an asynchronous mode at speed, ${ITR}_{letter}$, and ${ITR}_{com}$ (${p<0.05}$, Wilcoxon signed-rank using the Bonferroni correction).

\begin{table}[]
\caption{Performance of proposed CNN model using the similar input modalities used in previous works.}
\label{tab:first}
\begin{tabular}{lcll}
\hline
\multicolumn{1}{c}{Modality} & AUC (\%) & \multicolumn{1}{c}{Method} & \multicolumn{1}{c}{Dataset} \\ \hline
\cite{20} IR sensors & 80 & Appearance-based & 7094 images \\
\cite{19} Webcam & 94.39 & Appearance-based & MPIIGaze \\
\cite{14} Webcam & 88 & Appearance-based & 6000 images \\
\cite{18} Webcam & 84 & Appearance-based & 6800 images \\
\cite{17} Webcam & 98.77 & \begin{tabular}[c]{@{}l@{}}Appearance-based\end{tabular} & 75000 images \\
Proposed Webcam & 99.64 & Appearance-based & 2700 images \\ \hline
\end{tabular}
\end{table}



\begin{table}
\centering
\caption{Typing performance (mean and standard deviation across participants) for the mouse, eye-tracker, and webcam}
\label{tab:second}
\begin{tabularx}{\columnwidth}{@{}lXXX@{}}
\toprule
Conditions & \text{Speed (TER)} (letters/min) & \text{${ITR}_{letter}$} (bits/min) & \text{${ITR}_{com}$} (bits/min) \\
\midrule
Mouse & 18.3 ± 5.31 & 106.28 ± 30.85 & 118.45 ± 32.89 \\
Eye-tracker (Sy) & 12.60 ± 2.29 & 73.18 ± 13.28 & 88.66 ± 11.41 \\
Webcam (Sy) & 10.94 ± 1.89 & 63.56 ± 11 & 80.29 ± 15.72 \\
Eye-tracker (As) & 11.15 ± 2.90 & 64.79 ± 16.85 & 84.07 ± 17.30 \\
Webcam (As) & 7.86 ± 1.69 & 45.61 ± 9.81 & 71.62 ± 22.33 \\
\bottomrule
\end{tabularx}
\end{table}

\begin{figure}[]
  \includegraphics[width=0.45\textwidth]{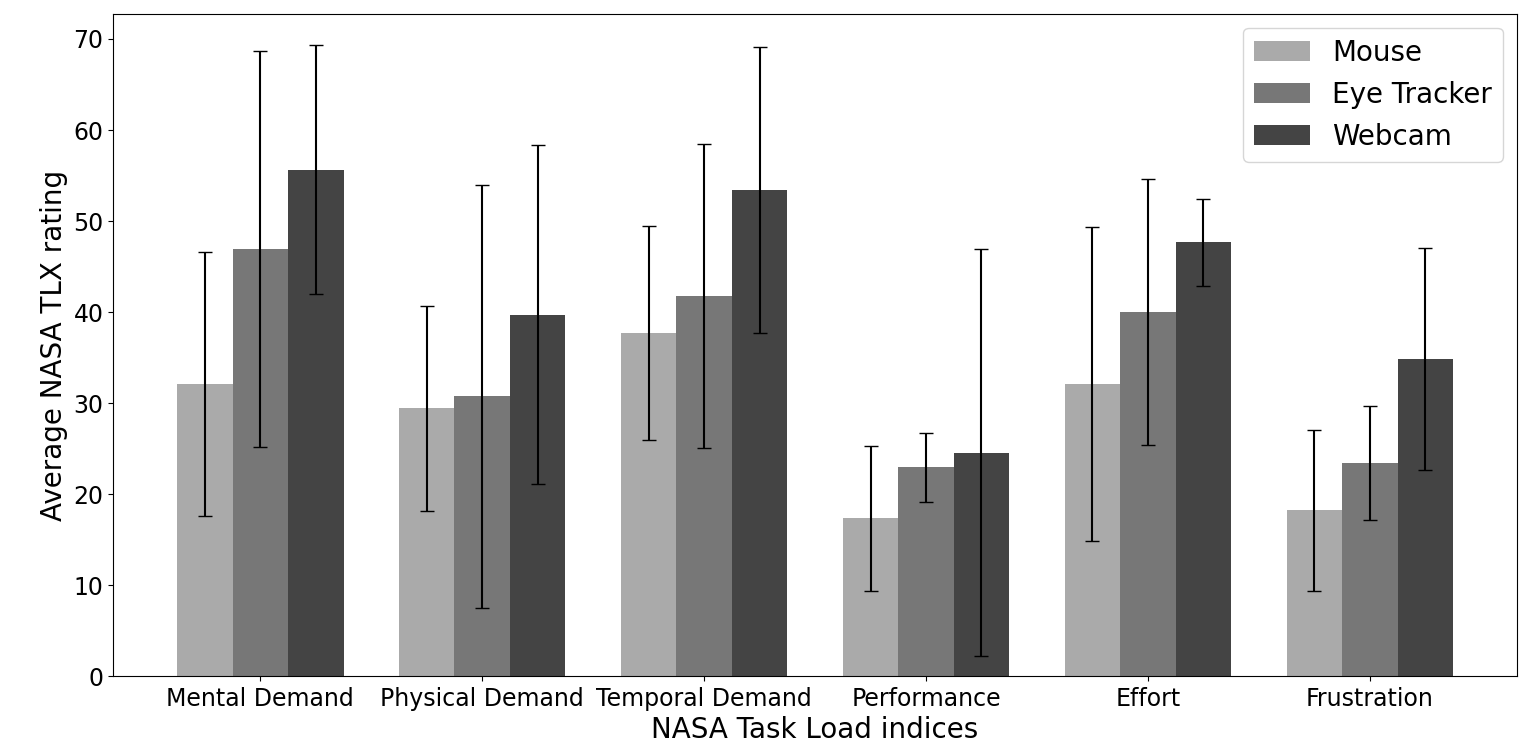}
  \caption{The average NASA TLX rating score across different input modalities. Standard errors across the different experimental conditions performed by participants for the respective input modality are indicated by the error bars.}
  \label{fig:nasa}
\end{figure}

\begin{figure}[ht!]
  
  \includegraphics[width=0.45\textwidth]{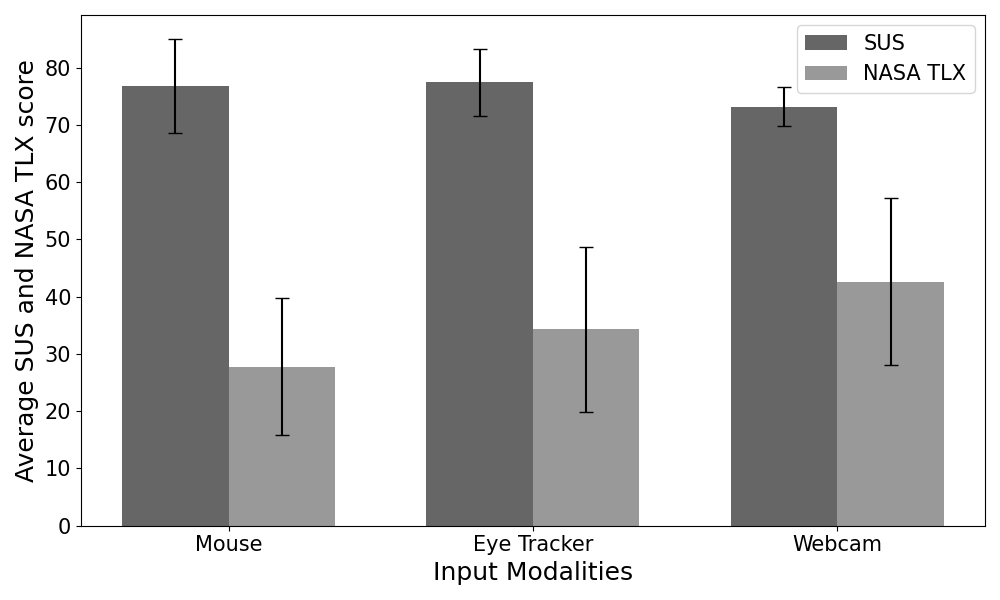}
  \caption{Average system usability scale (SUS) and NASA Task load
index (NASA-TLX) score for each input modality across all participants. Standard errors across the different experimental conditions performed by participants for the respective input modality are indicated by the error bars. 
}
  \label{fig:sus}
\end{figure}

\subsection{Subjective evaluation} \label{subsec:subjective evaluation}

The NASA Task Load Index (NASA-TLX) is an assessment tool used to check the workload experienced by users while performing any task. This index is used extensively to check how the user is experiencing different aspects of workload like mental demand, physical demand, Temporal demand, effort, frustration, and performance \cite{9,25,26}. The users need to score from 0 to 100 for each index (mental demand, physical demand, etc.) of the workload after performing the task on the virtual keyboard. The lesser the NASA-TLX score, the better the performance. NASA-TLX tests were conducted for each experimental condition. The average score over all the participants for each input modality was calculated as represented in Fig ~\ref{fig:nasa}. Overall average score was less then 40 for all conditions. 

The system usability scale (SUS) consists of ten different types of scales, providing us with a detailed analysis of the usability of our system \cite{9}. The user needs to scale from 0 to 4 for each of the 10 items present in the usability test. This test helps get user feedback and analyse different factors associated with the system (e.g., whether the system is beneficial for them, whether the system is complex, or whether it is easy to use). This usability scale mainly focuses on evaluating the system on three vital aspects of usability: satisfaction, usability, and efficiency. The final scores of the SUS test are reported on a scale of 0 to 100 for each item. The higher the score, the higher the usability of the system. The test was conducted for each experimental condition, and the averages were calculated across all the participants for each input modality. The average SUS score for each input modality is presented in Fig ~\ref{fig:sus}. The average score for each input modality was above 73, showing a good usability scale of the system.

\section{Discussion} \label{subsec:Discussion and Conclusion}

This work presents a multimodal gaze direction classification system, introducing two command selection methods: synchronous and asynchronous. These methods are designed for webcam-based gaze interaction, commonly used in eye-tracking~\cite{9} and brain-computer interface systems~\cite{29}. We evaluated the effectiveness of these approaches using a newly designed virtual keyboard (VK) application that integrates multimodal capabilities. We compared the typing performance of the VK application using a webcam with that of traditional eye-trackers. The appearance-based model for eye direction classification achieved accuracy levels that were significantly closer to those of eye-trackers. Specifically, when using a laptop webcam, we achieved a typing speed of \(10.94 \pm 1.89\) letters per minute, which is comparable to the \(12.60 \pm 2.29\) letters per minute achieved with an eye-tracker. Although there is a slight difference in speed, the affordability and accessibility of the webcam-based system make it an attractive option, particularly for individuals with disabilities in low-resource settings. Additionally, by employing synthesised eye images and transfer learning, our model reduces the need to collect real eye samples for the dataset, achieving an impressive accuracy of over 99\% with only 2,700 images of calibration data.

In addition to the gaze intimidation model based on appearance and typing performance, we included subjective evaluation metrics to gain a better understanding of the input modalities and overall system usability. The NASA-TLX ratings (see Fig ~\ref{fig:nasa}) indicate that users perceive the performance of our webcam-based system to be comparable to that of the eye-tracker, with only a slight difference in the performance index. Furthermore, the average System Usability Scale (SUS) scores reported by users across different input modalities show minimal differences, suggesting that the system performs well in all areas of usability, as noted by the participants.


It is important to note that the users did not have a chin rest during the experiment, which increased the likelihood of head movement. This setup may have led to an improvement in speed for individuals with motor and speech disabilities. According to Table \ref{tab:second}, the asynchronous mode did not perform well with the webcam. Whenever the model made an incorrect prediction, the asynchronous algorithm would restart the dwell time from the beginning, resulting in wasted time and reduced efficiency. In contrast, when using an eye-tracker, there were significantly fewer instances of incorrect predictions, which enhanced performance. Although the system showed good performance in controlled conditions, its long-term usability and integration into daily routines have not yet been assessed. The inclusion of mouse input served only as a baseline, so direct comparisons with eye-trackers or classifiers are not meaningful at this stage. However, by integration of predictive approaches~\cite{31}, such comparisons could become relevant in more complex tasks that require balancing speed and accuracy, particularly when using a denser classification grid with additional rows and columns. Future longitudinal studies and real-world deployments are needed to evaluate user adaptability, fatigue, and system acceptance, with the goal of supporting sustained, independent communication and enhancing quality of life. 


Future work will focus on evaluating the system with a larger and more diverse sample, including both healthy individuals and those with motor and speech impairments. In particular, collecting feedback from individuals with such disabilities will offer valuable insights into the system’s practical efficacy. Additionally, subsequent studies will compare the model’s performance against existing commercial solutions, using the findings to improve the overall user experience for the target demographic. This line of work could also be extended to support the early detection of dyslexia in children and aid in the development of targeted intervention programs~\cite{32,33}.

\section{Conclusion} \label{subsec:Conclusion}

This work demonstrates significant advancements in eye direction classification using deep learning techniques. By integrating both synchronous and asynchronous modes for command selection in webcam-based gaze interaction, it improves user engagement and accessibility, addressing the challenges posed by involuntary eye movements and low accuracy. The development of a multimodal virtual keyboard application facilitates efficient communication and accommodates a wide range of inputs, ensuring a versatile typing experience. The proposed appearance-based gaze estimation model achieved accuracy comparable to that of an eye-tracker when using a webcam. In testing with the virtual keyboard application, the synchronous algorithm outperformed the asynchronous one. This work can be integrated with existing infrastructure devices, enhancing the long-term usability of the system and its integration into the daily activities of potential users in low-resource settings.

\noindent{Acknowledgement:} This study was supported by Anusandhan National Research Foundation (ANRF) grant ANRF/ECRG/2024/002814/ENS.

\bibliographystyle{unsrt}
\bibliography{references}

\end{document}